\documentclass[a4paper,11pt]{article}
\usepackage{jheppub}
\usepackage{makecell}
\usepackage{cancel}
\usepackage[title]{appendix}

\newcommand{\Br}{{\mathcal{B}}}

\newcommand{\DDbar}{\ensuremath{\overset{\scalebox{0.32}{$(\,$}\rule[0.55pt]{0.17cm}{0.4pt}\scalebox{0.32}{$\,)$}}{D}{}^0}}

\title{\boldmath A model independent method for measurement of $B^{\pm}$ and $B^0$ meson production fractions at $\Upsilon(4S)$}

\author[a,b]{M. Yasaveev,}
\author[a,b]{P. Pakhlov,}
\author[a,b]{N. Peters,}
\author[a,b]{A. Mufazalova}
\affiliation[a]{International Laboratory of Elementary Particle Physics, National Research University Higher School of Economics, 20 Myasnitskaya Street, Moscow 101000, Russia}
\affiliation[b]{Laboratory of Heavy Quarks and Leptons, P.N. Lebedev Physical Institute of RAS, 53 Leninskiy Prospekt, Moscow 119991, Russia}
\emailAdd{myasaveev@hse.ru}
\emailAdd{ppakhlov@hse.ru}
\emailAdd{npeters@hse.ru}
\emailAdd{mufazalova.a.o@hse.ru}
\date{\today}

\abstract{This paper presents a detailed description of a model-independent method for the direct measurement of the $\Upsilon(4S) \to B^+ B^-$ and $\Upsilon(4S) \to B^0 \bar{B}^0$ production fractions at $B$ factories. The method is based on counting single- and double-inclusive charmed meson production at the $\Upsilon(4S)$ resonance, providing statistical tagging of $B^0\bar{B}^0$ and $B^+B^-$ events. A feasibility study indicates that a precision comparable to the current world average can be achieved without making any underlying assumptions, as the calculations rely solely on event yields.}

\arxivnumber{}

\begin{document}
\maketitle
\flushbottom

\date{\today}

\section{Introduction}

Electron--positron collisions at the center-of-mass energy near the $\Upsilon(4S)$ resonance provide a clean environment for studies of $B$ mesons. Since the $\Upsilon(4S)$ mass lies only slightly above the threshold for open-bottom production, its decays are dominated by the two channels $\Upsilon(4S)\to B^+B^-$ and $\Upsilon(4S)\to B^0\bar{B}^0$. Decays to final states without $B$ mesons are strongly suppressed and their total probability is measured to be $(0.26_{-0.02}^{+1.25})\%$~\cite{hflav:2024ctg}. For this reason, data collected at the $\Upsilon(4S)$ resonance play a central role in the experimental $B$-physics program.

A key quantity for many measurements is the ratio of the production fractions of charged and neutral $B$ meson pairs,

\begin{equation}
\frac{f^{+-}}{f^{00}} \equiv 
\frac{\Gamma(\Upsilon(4S)\to B^+B^-)}{\Gamma(\Upsilon(4S)\to B^0\bar{B}^0)} .
\end{equation}
This ratio and the individual production fractions $f^{+-}$ and $f^{00}$ are required to convert observed event yields into absolute branching fractions of $B^+$ and $B^0$ mesons~\cite{Belle:0}. It therefore affects precision determinations of Cabibbo--Kobayashi--Maskawa matrix elements~\cite{Belle:1}, tests of flavor symmetries~\cite{Belle:2}, and searches for physics beyond the Standard Model~\cite{Belle:3}. As the experimental precision of many $B$ decay measurements has improved significantly, the uncertainty on $f^{+-}/f^{00}$ has become an important limiting factor.

The $f^{+-}/f^{00}$ ratio is not expected to be exactly equal to unity. Small differences arise from the mass difference between charged and neutral $B$ mesons, which leads to different available phase space, as well as from electromagnetic effects and isospin-violating contributions in the strong interaction~\cite{Voloshin:2003gm}. Since the available phase space in $\Upsilon(4S)$ decays is small, these effects can be enhanced. Theoretical estimates of their size vary widely, which makes direct experimental measurements of $f^{+-}/f^{00}$ necessary~\cite{Bondar:2022kxv}.

Several measurements of $f^{+-}/f^{00}$ have been performed by the CLEO, BaBar, and Belle collaborations using different experimental methods. The most precise determination to date $f^{+-}/f^{00} = 1.065 \pm 0.012 \pm 0.019 \pm 0.047$~\cite{Belle:2022hka} is obtained using $B\to J/\psi K$ decays. Similar measurements were also performed at CLEO~\cite{CLEO:2000nea} and BaBar~\cite{BaBar:2004htr}. This approach is based on comparing the yields of charged and neutral $B$ meson decays and assumes that the corresponding partial decay widths are equal. Under this assumption, the observed difference in event yields arises only from the different production fractions, the difference between the $B^+$ and $B^0$ lifetimes, and the different reconstruction efficiencies for charged and neutral kaons. The dominant uncertainty of this method arises from the model-dependent assumption of isospin symmetry, which is measured to have an accuracy of $\sim 1\%$~\cite{LHCb:2025jva}.

Another class of measurements uses semileptonic~\cite{CLEO:2002leu} or dilepton~\cite{Belle:2002lms} events, where the ratio $f^{+-}/f^{00}$ is extracted from samples of $B\to \ell X$ decays. These analyses typically assume that the inclusive semileptonic decay widths of charged and neutral $B$ mesons are equal. While this assumption is expected to hold to good accuracy, it still relies on theoretical arguments based on heavy-quark expansion and suppressed isospin-breaking effects.

There is also a BaBar measurement of $f^{00} = 0.487 \pm 0.010 \pm 0.008$~\cite{BaBar:2005uwr}, obtained by comparing the number of events with one and with two partially reconstructed $B^0 \to D^{*-} \ell^+ \nu_{\ell}$ decays. This method is less sensitive to assumptions about isospin symmetry, as it relies on double-tagged semileptonic decays. However, it requires an estimate of the contribution from $D^{**}$ states, which are not measured directly and introduce an additional source of systematic uncertainty. The conceptual features of this approach are closely related to the strategy developed in this work and are discussed in more detail in Sec.~\ref{sec:babar}.


The current world-average values of $f^{+-} = 0.5113^{+0.0073}_{-0.0108}$, $f^{00} = 0.4861^{+0.0074}_{-0.0080}$, and $f^{+-}/f^{00} = 1.052 \pm 0.031$~\cite{hflav:2024ctg} combine results from CLEO, BaBar, and Belle. However, this averaging procedure requires particular care as the $f^{+-}/f^{00}$ ratio depends on the center-of-mass energy as shown in Ref~[Mizuk, Bondar]. The data sets used by different experiments were collected at slightly different center-of-mass energies and with different beam-energy spreads. These differences can introduce additional systematic effects that are difficult to quantify and are not fully accounted for in existing averages. This further limits the robustness of the combined result.

Thus, existing measurements of the $f^{+-}$ and $f^{00}$ production fractions are affected by model-dependent assumptions, limited experimental precision, and additional systematic uncertainties related to the operating conditions of the experiments. Therefore, a more precise and less model-dependent determination of this ratio is needed, as suggested in~\cite{Bernlochner:2023bad}. In this paper, we present a new method for measurement of the $f^{+-}$ and $f^{00}$ in $e^+e^-$ collisions at the $\Upsilon(4S)$ resonance.
The proposed method is model-independent, does not require knowledge of reconstruction or selection efficiencies, and relies on combinatorial counting without the need for kinematic distributions fit. The only background to be subtracted is the continuum contribution, which is also determined by counting. 


\section{Method\label{sec:theory}}

The method proposed in this work is conceptually based on the double-tag technique in $\Upsilon(4S)$ decays.  At the $\Upsilon(4S)$ resonance, $B$ mesons are produced in pairs, and the numbers of events with one or with two reconstructed decay signatures can be related to the underlying production fractions of charged and neutral $B$ mesons. By counting single-tag and double-tag yields, one can construct a system of equations that connects observable event rates to the $B$-meson production fractions. In the present work, this idea is generalized to a statistical tagging framework using inclusive decay signatures rather than a single exclusive channel. Before introducing the full formalism, we briefly review the implementation of the double-tag method in previous measurements, which serves as the conceptual starting point of our approach.

\subsection{Double-tag approach in $\Upsilon(4S)$ decays}
\label{sec:babar}
The double-tag strategy was implemented by the BaBar collaboration in the measurement of the neutral $B$-meson production fraction~\cite{BaBar:2005uwr}. In that analysis, $f^{00}$ was determined using partially reconstructed semileptonic decays $B^0 \to D^{*-} \ell^+ \nu_{\ell}$. Events with one reconstructed decay (single-tag events) and events with two reconstructed decays in the same event (double-tag events) were extracted from a fit to the missing-mass-squared distribution.

Denoting the product of the  $B^0 \to D^{*-} \ell^+ \nu_{\ell}$ branching fraction and the corresponding reconstruction efficiency as $\Br_{D^* \ell \nu}$, the expected numbers of single- and double-tag events can be written as
\begin{align}
N_{ST} &= 2 N_{\Upsilon(4S)} f^{00} \cdot\mathcal{B}_{D^*\ell\nu} , \\
N_{DT} &= N_{\Upsilon(4S)} f^{00} \cdot \mathcal{B}_{D^*\ell\nu}^2 .
\end{align}
Thus, the ratio of single- and double-tag yields allows the extraction of $f^{00}$ independently of the absolute value of the branching fraction:
\begin{equation}
f^{00} = \frac{N_{ST}^2}{4 N_{\Upsilon(4S)} N_{DT}}.
\end{equation}

Despite its apparent simplicity, this method is subject to several important limitations. A key issue is the possible contribution of $D^{*-}\ell^+$ pairs originating from charged $B$ decays. In particular, decays of the type $B^+ \to \bar{D}^{**0} \ell^+ \nu_{\ell}$, followed by $\bar{D}^{**0} \to D^{*-} \pi^+$, can produce final states that mimic the signal topology. Although the kinematic distributions of such processes differ from those of the signal decay $B^0 \to D^{*-} \ell^+ \nu_{\ell}$, a fraction of these events can populate the signal region due to finite detector resolution and the partial reconstruction technique. As a result, the extraction of $f^{00}$ requires modelling of the $B^+ \to D^{**0} \ell \nu$ contribution. The normalization and kinematic shape of this background component introduce non-negligible systematic uncertainties. 
In particular, exclusive semileptonic decays of charged $B$ mesons to excited charm states have not been fully measured experimentally; furthermore, neither the form factors nor the $D^{**}$ polarizations for such decays are well known. These factors limit the precision of the constraints on this contribution.

These considerations motivate a more general statistical approach. 
In contrast to the BaBar implementation, the present method does not attempt to isolate a pure $B^0$-initiated sample by suppressing contributions from  charged $B$ decays. Instead, the analysis exploits the fact that different reconstructed charm mesons are produced with different probabilities in $B^0$ and $B^+$ decays. In particular, the inclusive production rates of particles such as $D^0$ and $D^{*+}$ differ significantly between neutral and charged  $B$ mesons. As a result, the observed yields of these particles contain statistical information on the relative contributions of $B^0$ and $B^+$ decays in the data sample. Rather than treating the contribution of one $B$ species as a background to the other, both are incorporated explicitly into a system of equations relating the observed single- and double-tag yields to the underlying production fractions.

Consequently, the method does not rely on modelling kinematic distributions of excited charm states or on separating signal from signal-like components. The extraction of the production fractions is based only on counting reconstructed charm mesons and their combinations, making the approach largely insensitive to the detailed decay dynamics of intermediate states.

\subsection{Simplified system with semileptonic tags}

To illustrate the key idea of the method, we first consider a simplified scenario based on single- and double-tags using inclusive $D^0$ and $D^{*+}$ signatures. We selected the $D^0$ and $D^{*+}$ mesons due to their relatively high reconstruction efficiency and signal purity. In double-tag events the two reconstructed charm mesons are required to originate from different $B$ mesons. As will be discussed below, this condition can be achieved experimentally by imposing suitable kinematic requirements.

In this simplified system, we do not distinguish between right- and wrong-sign charge combinations. Instead, the observables $D^0$ and $D^{*+}$ are understood as inclusive sums over the corresponding charge-conjugate final states. Neutral $B^0$--$\bar{B}^0$ mixing is therefore implicitly included in the effective inclusive rates, and the production fractions satisfy the normalization condition
\begin{equation}
f^{+-} + f^{00} = 1.
\end{equation}

We define the effective inclusive branching fractions
\[
\mathcal{B}(B^{0/+} \to D^{*+} X) \equiv \Br^{0/+}_{+},
\qquad
\mathcal{B}(B^{0/+} \to D^{0} X) \equiv \Br^{0/+}_{0},
\]
where $\Br^{\alpha}_{i}$ denotes the product of the physical branching fraction and the corresponding reconstruction efficiency.

Since $D^{*\pm}$ mesons originate predominantly from neutral $B$ decays, while $D^0/\bar{D}^0$ mesons are produced more frequently in charged $B$ decays, their yields provide statistical sensitivity to the production fractions.

The expected numbers of single-tag events are
\begin{align}
N_{D^{*+}} &= 2 N_{\Upsilon(4S)} \left(f^{00}\,\Br^0_{+} + (1-f^{00})\,\Br^+_{+}\right), \\
\label{ST}
N_{D^0} &= 2 N_{\Upsilon(4S)} \left(f^{00}\,\Br^0_{0} + (1-f^{00})\,\Br^+_{0}\right),
\end{align}
where the factor of two accounts for the two $B$ mesons produced in each event.

These two equations depend on five unknown parameters: the four inclusive branching fractions and the production fraction $f^{00}$. To obtain additional constraints, we introduce double-tag yields corresponding to events in which charm mesons are reconstructed on both sides:

\begin{align}
N_{D^{*+}D^{*-}} &= N_{\Upsilon(4S)} \left( f^{00}\,(\Br^0_{+})^2 + (1-f^{00})\,(\Br^+_{+})^2 \right), \\
N_{D^{*+} \bar{D}^0} &= 2N_{\Upsilon(4S)} \left(f^{00}\,\Br^0_{+}\Br^0_{0} + (1-f^{00})\,\Br^+_{+}\Br^+_{0} \right),\\
N_{D^0\bar{D}^0} &= N_{\Upsilon(4S)} \left( f^{00}\,(\Br^0_{0})^2
+ (1-f^{00})\,(\Br^+_{0})^2 \right).
\label{DT}
\end{align}

At first sight, the inclusion of double-tag observables appears to provide sufficient information to determine all unknown parameters, since the system consists of five equations for five unknowns. However, the observables are not independent. In fact, the mixed double-tag yield is constrained by the single-tag and same-species double-tag yields. This relation can be written explicitly as
\begin{equation}
N_{D^{*+} \bar{D}^0} =
N_{D^{*+}}N_{D^0}
- 2\sqrt{\left(N_{D^{*+}D^{*-}} - \Big(\frac{1}{2}N_{D^{*+}}\Big)^2\right)
\left(N_{D^{0}\bar{D}^0} - \Big(\frac{1}{2}N_{D^{0}}\Big)^2\right)} .
\end{equation}

The origin of this degeneracy becomes clear if the system is interpreted as a mixture of two components. Once the first moments and the variances are fixed, the mixed second moment is determined automatically. In this sense, the double-tag observables do not provide an additional independent constraint. The system therefore remains degenerate and cannot be used to extract all unknown parameters simultaneously.

This degeneracy is not specific to the particular choice of $D^0$ and $D^{*+}$ tags, but is a generic feature of systems based solely on inclusive single- and double-tag observables. Adding additional particle species does not remove it: each new tag introduces two additional effective branching fractions (for $B^0$ and $B^+$ decays), while providing only the corresponding single-tag yield and the double-tag yield with itself. As a result, the extended system remains algebraically degenerate.

\subsection{Lifting the degeneracy through charge correlations}
\label{sec:mixing_idea}

In the simplified system described above, only charge-averaged observables are considered. In this case, all double-tag yields retain a factorized structure, which leads to the algebraic degeneracy. A qualitatively new ingredient is provided by $B^0$–$\bar B^0$ mixing. While charged $B$ mesons do not oscillate, neutral mesons mix with probability $\chi_d = 0.186 \pm 0.001$~\cite{PDG:2024cfk}. Mixing modifies the double-tag observables and breaks their simple factorized structure.

To exploit this effect, it is necessary to resolve the charge of the reconstructed particles. This requires distinguishing between right- and wrong-sign $D$ mesons, which effectively doubles the number of inclusive branching fractions entering the system. As a consequence, the number of unknown parameters increases to nine: eight charge-specific effective branching fractions and the production fraction $f^{00}$. The number of double-tag observables also increases due to charge separation; however, the system still provides only eight independent equations. Therefore, even though the algebraic degeneracy of the simplified system is lifted, the charge-resolved system remains underconstrained. Additional independent observables are thus required.

A natural extension is to include charged lepton signatures. Since the inclusive semileptonic branching fractions are similar for $B^0$ and $B^+$ mesons, the presence of a lepton alone does not distinguish between $B^0$ and $B^+$ decays and therefore does not provide direct sensitivity to the ratio $f^{+-}/f^{00}$. However, a positively charged lepton tags the decay of a $B$ meson and allows the relative rates of right-sign and wrong-sign $D$ meson production to be constrained. This tagging is statistical, as it is diluted by secondary leptons; the dilution can be accounted for by introducing the branching fraction for wrong-sign lepton production, $\mathcal{B}(B \to \ell^- X)$.

In addition, the inclusion of leptons enables the use of $B \to D^{(*)}\ell X$ decays as inclusive tags. These tags play a role similar to that in the BaBar analysis, but are incorporated here at the level of inclusive counting observables. As a result, the approach avoids the model-dependent assumptions of the BaBar method: $D^{(*)}\ell$ signatures are allowed to originate from decays of both $B$ mesons, and no use is made of kinematic distributions in the extraction of the yields.

\subsection{Complete charge-resolved system}
\label{sec:real_system}

We now construct the complete system of observables incorporating charge separation and all relevant inclusive contributions. As discussed above, the key ingredients of the method are three types of reconstructed particles: 
$D^0/\bar D^0$, $D^{*\pm}$, and $\ell^\pm$. 
The combinations of these particles are interpreted as single- and double-tag signatures, and the corresponding observables are expressed in terms of inclusive branching fractions of the form
$B \to D X$, 
$B \to \ell X$, 
and $B \to D \ell X$.

In principle, charge separation should be implemented for all such modes, requiring independent branching fractions for right- and wrong-sign combinations such as  $B \to \bar{D}\ell^+ X$ and $B \to D\ell^- X$. However, the doubly wrong-sign semileptonic transitions, $B \to D \ell^- X$, are strongly suppressed and the present method does not provide sufficient sensitivity to determine them reliably. To avoid introducing poorly constrained tiny parameters that the fit might find non-physically negative, semileptonic $D\ell$ combinations are therefore treated inclusively in charge. This reduces the number of unknown branching fractions while retaining charge resolution for single $D$ and $\ell$ tags, which is sufficient to break the factorized structure of part of the system.

Double-tag observables are additionally affected by decays of the type $B \to D \bar D X$, which produce two charm mesons in a single decay. Such contributions would introduce additional independent branching fractions and significantly enlarge the parameter space. However, they can be completely suppressed by imposing a requirement on the $D$-meson momentum. Since the $B$ mesons are produced nearly at rest in the $\Upsilon(4S)$ frame, two energetic charm mesons cannot be emitted into the same hemisphere. A suitable kinematic selection therefore effectively removes these configurations from the sample.

In contrast, dilepton decays $B \to \ell^+\ell^- X$ are retained. Although they introduce additional branching fractions, vetoing them would substantially reduce the statistical power of the method.

With this setup, the unknown parameters consist of the production fractions $f^{+-}$ and $f^{00}$ and the following inclusive branching fractions, defined within the imposed kinematic selection:
\[
\mathcal{B}(B^{0/+} \to D^{*-} X), \quad
\mathcal{B}(B^{0/+} \to \bar{D}^0 X), \quad
\mathcal{B}(B^{0/+} \to \ell^+ X),
\]
\[
\mathcal{B}(B^{0/+} \to D^{*+} X), \quad
\mathcal{B}(B^{0/+} \to D^0 X), \quad
\mathcal{B}(B^{0/+} \to \ell^- X),
\]
\[
\mathcal{B}(B^{0/+} \to D^{*\mp}\ell^{\pm} X), \quad
\mathcal{B}(B^{0/+} \to \DDbar \ell^{\pm} X), \quad
\mathcal{B}(B^{0/+} \to \ell^+\ell^- X).
\]

The system contains eighteen effective branching fractions and two production fractions, $f^{+-}$ and $f^{00}$, for a total of twenty unknown parameters. The corresponding observables are the charge-resolved single- and double-tag yields constructed from these particles. The number of equations exceeds the number of unknown parameters, and the resulting system is algebraically non-degenerate.

\section{Selection criteria and analysis strategy}

The selection strategy exploits the kinematic properties of $B$ mesons produced at the $\Upsilon(4S)$ resonance. In the center-of-mass (CMS) frame, the $B$ mesons are produced nearly at rest, with an energy $E_B \simeq 5.29~\mathrm{GeV}$ and a momentum $p_B \simeq 0.34~\mathrm{GeV}/c$. As a consequence, the decay products of a single $B$ meson are kinematically constrained to have a limited total momentum in this frame.

This feature can be used to suppress configurations in which two charm mesons originate from the decay of a single $B$ meson and are emitted into the same hemisphere. If two $D$ mesons are emitted into the same hemisphere and each carries a CMS momentum exceeding $1.0~\mathrm{GeV}/c$, their combined momentum necessarily exceeds $\sqrt{2}\times 1.0 \simeq 1.4~\mathrm{GeV}/c$. Momentum conservation therefore requires the remaining decay system to carry at least $1.4 - 0.34 \simeq 1.1~\mathrm{GeV}/c$ in the opposite direction. Even in the most favorable configuration, where the remaining system is treated as a single massless particle, its energy would then have to exceed $1.1~\mathrm{GeV}$. 

The total energy of the decay products would therefore satisfy
\[
E_{\rm tot} > 2\sqrt{m_D^2 + (1.0~\mathrm{GeV})^2} + 1.1~\mathrm{GeV} \simeq 5.3~\mathrm{GeV},
\]
which exceeds the available $B$-meson energy $E_B \simeq 5.29~\mathrm{GeV}$. Such configurations are therefore kinematically forbidden. Taking into account the beam-energy spread of about $5~\mathrm{MeV}$, the requirement $p_D \gtrsim 1.05~\mathrm{GeV}/c$ provides a robust suppression of $B \to D\bar D X$ decays in which both charm mesons are emitted into the same hemisphere. 

In contrast, for $D$ mesons originating from the decays of two different $B$ mesons, no strong angular correlation is expected in the $\Upsilon(4S)$ frame, and approximately $50\%$ of such events populate the same-hemisphere region.

For lepton selection, $B$-factory analyses typically impose a requirement of $p_\ell > 1.0~\mathrm{GeV}/c$ in the CMS frame for reliable identification. A more restrictive requirement, $p_\ell \gtrsim 1.7~\mathrm{GeV}/c$, would allow kinematic suppression of semileptonic decays of the type $B \to D \ell \nu$ and $B \to \ell^+ \ell^- X$, but such a cut significantly reduces the signal efficiency. Therefore, in the present analysis, the standard requirement $p_\ell > 1.0~\mathrm{GeV}/c$ is retained.

The yields of single- and double-tag events containing $D$ mesons are determined using one- and two-dimensional fits to the invariant-mass distributions. The $D^{*+}$ meson is reconstructed through the decay $D^{*+} \to D^0 \pi^+$, and $D^0$ through the $K^- \pi^+$ decay.

Since the method relies on effective inclusive rates rather than physical branching fractions, additional selection requirements can be applied without introducing a bias in the determination of the production fractions. In particular, kinematic variables such as the $D^0$ decay angle — defined as the angle between the kaon momentum and the boost direction of the laboratory frame in the $D^0$ rest frame — can be used to further suppress background contributions. In a realistic analysis, the choice of such requirements is expected to be optimized to achieve the best statistical sensitivity.

\section{Notations and definitions}
\label{notations}

We use the following notation throughout the paper. The types of $B$ mesons are denoted by a Greek index, $B^{\alpha}$ with $\alpha \in \{0,\bar{0},+,-\}$, corresponding to $B^0$, $\bar{B}^0$, $B^+$, and $B^-$, respectively. Reconstructed daughter particles are denoted by the symbol $d_i$, where the index $i$ labels the particle species and its charge/flavor. In this work we consider
\[
d_{\pm 1} \equiv D^{*\mp}, \qquad
d_{\pm 2} \equiv \DDbar, \qquad
d_{\pm 3} \equiv \ell^{\pm}.
\]

We define the inclusive branching fractions
\[
\mathcal{B}(B^{\alpha} \to d_i X) \equiv \mathcal{B}^{\alpha}_i .
\]
For example, $\mathcal{B}(B^0 \to \ell^+ X) \equiv \mathcal{B}^{0}_{3}$ and $\mathcal{B}(B^+ \to D^0 X) \equiv \mathcal{B}^{+}_{-2}$. In addition, double-inclusive branching fractions are defined as
\[
\mathcal{B}(B^{\alpha} \to d_i d_j X) \equiv \mathcal{B}^{\alpha}_{ij},
\]
which account for processes such as $B \to D^{*\mp}\ell^{\pm} X$ ($\mathcal{B}^{\alpha}_{13}$) and 
$B \to \ell^+\ell^- X$ ($\mathcal{B}^{\alpha}_{33}$).

Throughout this paper, $CP$ violation is neglected, so that $\mathcal{B}^{0/+}_{i} = \mathcal{B}^{\bar{0}/-}_{-i}$. Charge-conjugate processes are implied throughout. Accordingly, when counting single- and double-tag yields, only positive indices are used, with
\[
N_{d_i} \equiv N_{d_i} + N_{d_{-i}}, 
\qquad
N_{d_i d_j} \equiv N_{d_i d_j} + N_{d_{-i} d_{-j}},
\qquad
N_{d_i d_{-j}} \equiv N_{d_i d_{-j}} + N_{d_{-i} d_j}.
\]

Configurations involving three or four reconstructed daughter particles are considered only in cases where at least one or two leptons are present, respectively. Such observables are denoted as
\[
N_{d_i d_j \ell}
\quad \text{and} \quad
N_{d_i d_j \ell \ell}.
\]

\section{System of equations}
\label{equations}

In this section, we construct the system of equations relating the measured event yields to the inclusive and double-inclusive branching fractions defined in Sec.~\ref{notations}. The observables are grouped according to the number of reconstructed daughter particles. For brevity, the equations in this section are written in a compact form. The complete set of expressions for all observables is given in Appendix~\ref{appendixA}.

\subsection{Single-daughter yields}

The expected yield of inclusive single-daughter production in $\Upsilon(4S)$ decays can be written as
\begin{eqnarray}
\label{eq:1}
\begin{split}
N_{d_i} = 2N_{\Upsilon(4S)} \Big[ f^{00}\big(\mathcal{B}^0_i + \mathcal{B}^{\bar{0}}_i\big)
+ f^{+-}\big(\mathcal{B}^+_i + \mathcal{B}^-_i\big)\Big].
\end{split}
\end{eqnarray}
This expression provides three equations, corresponding to the three particle species considered.

\subsection{Double-daughter yields}

The expected number of pairs of daughter particles of the same charge is
\begin{eqnarray}
\label{eq:2}
\begin{split}
N_{d_i d_j} = C_2^{i+j-2}\times N_{\Upsilon(4S)} \Bigg[ f^{00} \Big( (1-\chi_d)\big(\mathcal{B}^0_i \mathcal{B}^{\bar{0}}_j + \mathcal{B}^0_j \mathcal{B}^{\bar{0}}_i\big)
+ \chi_d \big(\mathcal{B}^0_i \mathcal{B}^0_j + \mathcal{B}^{\bar{0}}_i \mathcal{B}^{\bar{0}}_j\big) \Big) \\
+ f^{+-} \big(\mathcal{B}^+_i \mathcal{B}^-_j + \mathcal{B}^+_j \mathcal{B}^-_i\big) \Bigg],
\end{split}
\end{eqnarray}
where $\chi_d = (18.60 \pm 0.11)\%$ is the time-integrated $B^0$--$\bar{B}^0$ mixing probability, and the combinatorial factor $C_2^{i+j-2}$ gives the number of permutations (1 for $i = j$ and 2 for $i \neq j$). This expression adds six equations to the system.

The expected number of opposite-sign $D$ mesons ($i, j = 1,2$) emitted into the same hemisphere is given by
\begin{eqnarray}
\label{eq:4}
\begin{split}
N_{d_i d_{-j}} = \frac{C_2^{i+j-2}}{2}\times N_{\Upsilon(4S)} \Bigg[f^{00} \Big( (1-\chi_d)\big(\mathcal{B}^0_i \mathcal{B}^{\bar{0}}_{-j} 
+ \mathcal{B}^0_{-j} \mathcal{B}^{\bar{0}}_i\big)
+ \chi_d \big(\mathcal{B}^0_i \mathcal{B}^0_{-j} + \mathcal{B}^{\bar{0}}_i \mathcal{B}^{\bar{0}}_{-j}\big)\Big) \\
+ f^{+-} \big(\mathcal{B}^+_i \mathcal{B}^-_{-j} + \mathcal{B}^+_{-j} \mathcal{B}^-_i\big) \Bigg],
\end{split}
\end{eqnarray}
where the factor $1/2$ accounts for the efficiency of the hemisphere requirement. This contributes three independent equations.

For pairs involving leptons, the expected yields are
\begin{eqnarray}
\label{eq:5}
\begin{split}
N_{d_i \ell^-} = C_2^{i+j-2} \times N_{\Upsilon(4S)} \Big[f^{00} \Big( (1-\chi_d)\big(\mathcal{B}^0_i \mathcal{B}^{\bar{0}}_{-3} 
+ \mathcal{B}^0_{-3} \mathcal{B}^{\bar{0}}_i\big)
+ {\chi_d} \big(\mathcal{B}^0_i \mathcal{B}^0_{-3} + \mathcal{B}^{\bar{0}}_i \mathcal{B}^{\bar{0}}_{-3}\big)\Big) \\
+ f^{+-} \big(\mathcal{B}^+_i \mathcal{B}^-_{-3} + \mathcal{B}^+_{-3} \mathcal{B}^-_i\big)\Big]  
+ 2N_{\Upsilon(4S)}\Big[f^{00}\mathcal{B}^0_{i3} + f^{+-}\mathcal{B}^+_{i3}\Big],
\end{split}
\end{eqnarray}
where the contribution of pairs originating from a single $B$ meson is explicitly included. This provides three additional equations.

\subsection{Triple-daughter yields}

For configurations with two $D$ mesons, a requirement is imposed that they are emitted into the same hemisphere. This requirement introduces a factor of $1/2$, which we denote by $\alpha_{ij}$. It is defined as $\alpha_{ij} = 1/2$ for $(i,j) \in \{1,2\} \times \{1,2\}$ and $\alpha_{ij} = 1$ otherwise. The expected yields are then
\begin{eqnarray}
\label{eq:8}
\begin{split}
N_{d_i d_j \ell} = \alpha_{ij} \times C_2^{i+j-2}\times N_{\Upsilon(4S)} \Bigg[ f^{00} \Big(\big(\mathcal{B}^0_i + \mathcal{B}^0_{-i}\big)\mathcal{B}^0_{j3} + \big(\mathcal{B}^0_j + \mathcal{B}^0_{-j}\big)\mathcal{B}^0_{i3}\Big) \\
+ f^{+-} \Big(\big(\mathcal{B}^+_i + \mathcal{B}^+_{-i}\big)\mathcal{B}^+_{j3} + \big(\mathcal{B}^+_j + \mathcal{B}^+_{-j}\big)\mathcal{B}^+_{i3}\Big)\Bigg].
\end{split}
\end{eqnarray}
These equations contribute six relations.

\subsection{Four-daughter yields}

For configurations containing two $D$ mesons and two leptons:
\begin{eqnarray}
\label{eq:13}
\begin{split}
N_{d_i d_j \ell \ell} = \alpha_{ij}\times C_2^{i+j-2} \times N_{\Upsilon(4S)} \Big[ f^{00} \mathcal{B}^0_{i3}\mathcal{B}^0_{j3} + f^{+-} \mathcal{B}^+_{i3}\mathcal{B}^+_{j3}\Big].
\end{split}
\end{eqnarray}
These relations provide six equations.

In total, the full system consists of twenty-seven equations and depends on twenty free parameters. The unknown parameters are determined by minimizing a $\chi^2$ function constructed from the measured yields and their uncertainties:
\begin{eqnarray}
\label{eq:15}
\begin{split}
\chi^2 = \sum_{k = 1}^{27} \frac{\big(N_{k}^{\rm meas} - N_{k}\big)^2} {(\delta N_{k}^{\rm meas})^2}.
\end{split}
\end{eqnarray}

\subsection{External constraints and discrete ambiguity}

The extended system of equations defined above no longer exhibits a continuous degeneracy. However, we observe the presence of additional discrete solution of the fit. This solution arises from an interplay between $B^0$--$\bar{B}^0$ mixing and wrong-sign charm production. In particular, the appearance of same-sign charm configurations can be accommodated either through neutral-$B$ mixing or through an enhanced rate of wrong-sign decays. Since both effects contribute to the same set of observables, the fit may converge to alternative solutions in which these contributions are redistributed, leading to physically distinct but comparably good descriptions of the data.

To resolve this ambiguity and stabilize the fit, external constraints are imposed on a subset of the inclusive branching fractions. In particular, we constrain the following four branching fractions:
\[
\mathcal{B}(B^0 \to D^{*+} X), \quad
\mathcal{B}(B^0 \to D^{*-} X), \quad
\mathcal{B}(B^+ \to D^{*+} X), \quad
\mathcal{B}(B^+ \to D^{*-} X),
\]
which are known from independent experimental measurements. The central values $\Br_k^{\rm meas}$ are taken from Ref.~\cite{BaBar:2006wbf}, and a conservative relative uncertainty of $20\%$ is assigned to each of them.

We note, however, that the inclusion of all four constraints is not essential. In practice, it is sufficient to constrain only one class of charm branching fractions (e.g., either $B \to D^{*+} X$ or $B \to D^{*-} X$) to eliminate the additional solution. The use of the full set of constraints is therefore redundant and serves primarily to improve numerical stability.

These constraints are incorporated into the fit by adding penalty terms to the $\chi^2$ function:
\begin{equation}
\chi^2_{\rm constr} = \sum_{k=1}^{4}
\frac{\big(\Br_k - \Br_k^{\rm meas}\big)^2}
{\left(\delta \Br_k^{\rm meas}\right)^2}.
\end{equation}
The total $\chi^2$ minimized in the fit is then given by
\[
\chi^2_{\rm total} = \chi^2 + \chi^2_{\rm constr}.
\]

The role of these constraints is not to provide precise external input, but to exclude unphysical solutions in which the wrong-sign contributions are artificially enhanced. We have verified that the resulting determination of the production fractions $f^{+-}$ and $f^{00}$ is stable with respect to reasonable variations of the assumed uncertainties, indicating that the fit is driven primarily by the data rather than by the external inputs.

\section{Feasibility study}

We estimate the expected precision achievable at modern $B$-factory experiments using data collected at the $\Upsilon(4S)$ resonance and in the nearby continuum. In addition, the robustness of the extracted parameters is studied with respect to relevant systematic effects, including uncertainties in inclusive $D$-meson branching fractions, reconstruction efficiencies, and background subtraction.

\subsection{Monte-Carlo simulation}

Since the proposed method relies on counting reconstructed charm mesons and leptons, a toy Monte-Carlo simulation is employed. In this simulation, only the ancestor structure of the decay chain is generated, namely $\Upsilon(4S) \to B^0\bar{B}^0$, $B^+B^-$, or non-$B\bar{B}$ events, followed by $B \to d_i X$, $B \to d_i d_j X$, or $B \to \text{no } d_i$, where $d_i$ denotes a reconstructed daughter particle as defined in Sec.~\ref{notations}. The decay probabilities and reconstruction efficiencies used in the simulation are summarized in Table~\ref{tab:dec}.

All branching fractions quoted below, unless stated otherwise, are taken from Ref.~\cite{PDG:2024cfk}.

The fraction for $B^0 \to D^{*-} X$ is estimated from the measured rate 
$\mathcal{B}(B^0 \to D^- X) = (36.9 \pm 3.3)\%$, assuming that decays into 
$D^{*-}$ mesons are enhanced by a factor of two relative to those into $D^-$ mesons. The inclusive branching fraction $\mathcal{B}(B^0 \to D^- X)$ also contains a contribution from $B^0 \to D^{*-} X$ followed by $D^{*-} \to D^- X$, whose fraction is $(32.3 \pm 0.6)\%$. Taking this into account, we estimate the corresponding branching fraction $\mathcal{B}(B^0 \to D^{*-} X)$ to be approximately $45\%$.

The remaining $B \to D^{*\pm} X$ branching fractions, $\mathcal{B}(B^0 \to D^{*+} X)$, $\mathcal{B}(B^+ \to D^{*-} X)$, and $\mathcal{B}(B^+ \to D^{*+} X)$, are assumed to be close to the corresponding $B \to D^{\pm} X$ rates: $(9.9 \pm 1.2)\%$, $(2.5 \pm 0.5)\%$, and $(2.3 \pm 1.2)\%$, respectively. In these channels, the contribution from $D^{*+}$ decays via $D^{*+} \to D^{+} X$ is subtracted from the inclusive rates. The efficiency for reconstructing the soft pion from this decay is assumed to be $60\%$. Decays proceeding via $D^{*+}$ therefore contribute effectively to reconstructed $D^0$ final states with a factor of approximately $0.68 \times 0.4 \approx 0.27$, reflecting the limited reconstruction efficiency of the soft pion.

The inclusive branching fractions for $B \to D^0/\bar{D}^0 X$ are 
$\mathcal{B}(B^0 \to D^0 X) = (8.1 \pm 1.5)\%$, 
$\mathcal{B}(B^0 \to \bar{D}^0 X) = (47.4 \pm 2.8)\%$, 
$\mathcal{B}(B^+ \to D^0 X) = (8.6 \pm 0.7)\%$, and 
$\mathcal{B}(B^+ \to \bar{D}^0 X) = (79 \pm 4)\%$. 
The contribution from $D^{*+}$ decays is subtracted from these inclusive rates. The branching fraction for $D^0 \to K^- \pi^+$ is $(3.95 \pm 0.03)\%$. The efficiency of the requirement on the $D$ momentum in the center-of-mass frame, $p_{D} > 1.05~\mathrm{GeV}/c$, is estimated using the momentum distribution reported in Ref.~\cite{BaBar:2006wbf}. The tracking and particle identification efficiency for kaons and pions from $D^0$ decay is assumed to be $80\%$.

The branching fractions for $B \to \ell^{+} X$ decays are 
$\mathcal{B}(B^0 \to \ell^+ X) = (20.66 \pm 0.56)\%$ and 
$\mathcal{B}(B^+ \to \ell^+ X) = (21.98 \pm 0.56)\%$. 
The branching fractions for wrong-sign leptons are estimated from $\mathcal{B}(B \to D X)$ and $\mathcal{B}(D \to \ell^+ X)$, with 
$\mathcal{B}(D^0 \to \ell^+ X) = (13.3 \pm 0.6)\%$ and 
$\mathcal{B}(D^+ \to \ell^+ X) = (33.7 \pm 3.2)\%$. 
This leads to the estimates $\mathcal{B}(B^0 \to \ell^- X) \approx 19\%$ and $\mathcal{B}(B^+ \to \ell^- X) \approx 7\%$.

The efficiency of the requirement $p_{\ell} > 1.0~\mathrm{GeV}/c$ 
is taken from Ref.~\cite{Belle:2002avk}, and the lepton reconstruction efficiency is assumed to be $80\%$. The branching fraction for $B \to \ell^+ \ell^- X$ is estimated as the product of the corresponding single-inclusive branching fractions. The contribution from $B \to J/\psi X \to \ell^+ \ell^- X$ is neglected at this stage.

The branching fractions for $B^0$ and $B^+$ decays into $D \ell X$ are measured by BaBar to be close to $9\%$~\cite{BaBar:2007xlq}. Since the branching fractions for $B^0 \to D^- \ell^+ \bar{\nu}_\ell$ and $B^0 \to D^{*-} \ell^+ \bar{\nu}_\ell$ are $(2.10 \pm 0.07)\%$ and $(4.87 \pm 0.09)\%$, respectively, the branching fraction for $B^0 \to D^{*\mp} \ell^{\pm} X$ is estimated to be $5\%$. The remaining contribution to $D \ell X$ final states is assumed to be shared equally between $D^0$ and $D^+$; this assumption is sufficient for the feasibility study and does not affect the general conclusions. Accordingly, the branching fraction for $B^0 \to \DDbar \ell^{\pm} X$ is estimated to be $1\%$.

For charged $B$ mesons, the branching fractions for $B^+ \to \bar{D}^0 \ell^+ \bar{\nu}_\ell$ and $B^+ \to \bar{D}^{*0} \ell^+ \bar{\nu}_\ell$ are $(2.26 \pm 0.07)\%$ and $(5.26 \pm 0.10)\%$, respectively. The branching fraction for $B^+ \to \DDbar \ell^{\pm} X$ is therefore estimated to be $8\%$, while that for $B^+ \to D^{*\mp} \ell^{\pm} X$ is taken to be $0.5\%$. The efficiencies for $D \ell$ pairs are approximated as the product of the corresponding single-particle efficiencies; correlations between reconstruction efficiencies are neglected at this stage.

\begin{table}[ht]
\centering
\begin{tabular}{|c|c|c|c|}
\hline\hline
Particle & Decay mode & \makecell[c]{Branching \\ fraction (\%)} & Efficiency (\%) \\ \hline\hline
$\Upsilon(4S)$ & $B^0\bar{B}^0$ & $f^{00}(1-\chi_d)$ & 100 \\
 & $B^0 B^0$ & $f^{00}\chi_d/2$ & 100 \\
 & $\bar{B}^0 \bar{B}^0$ & $f^{00}\chi_d/2$ & 100 \\
 & $B^+ B^-$ & $f^{+-}$ & 100 \\
 & non-$B\bar{B}$ & $1-f^{00}-f^{+-}$ & 100 \\ \hline
 $B^+$ & $D^{*+} X$ & 2 & 35 \\
 & $D^{*-} X$ & 2 & 45 \\
 & $D^0 X$ & 8 & 20 \\
 & $\bar{D}^0 X$ & 78 & 70 \\
 & $\ell^+ X$ & 22 & 65 \\
 & $\ell^- X$ & 7 & 3 \\
 & $D^{*\mp} \ell^{\pm} X$ & 0.5 & 30\\
 & $\DDbar \ell^{\pm} X$ & 8 & 45\\
 & $\ell^+ \ell^- X$ & 1.5 & 2\\ \hline
$B^0$ & $D^{*+} X$ & 7.5 & 50 \\
 & $D^{*-} X$ & 45 & 80 \\
 & $D^0 X$ & 5 & 20 \\
 & $\bar{D}^0 X$ & 29 & 65 \\
 & $\ell^+ X$ & 21 & 65 \\
 & $\ell^- X$ & 19 & 3 \\
 & $D^{*\mp} \ell^{\pm} X$ & 5 & 50\\
 & $\DDbar \ell^{\pm} X$ & 2.5 & 40\\
 & $\ell^+ \ell^- X$ & 4 & 2\\ \hline
$D^{*+}$ & $D^0 \pi^+$ & 67.7 & 60 \\
$D^0$ & $K^- \pi^+$ & 3.95 & 64 \\ \hline\hline
\end{tabular}
\caption{
Decay chain and input probabilities used in the toy Monte-Carlo simulation.}
\label{tab:dec}
\end{table}

\subsection{Fit to the system of equations}

Multiple pseudo-experiments are generated, each consisting of $10^9$ events, corresponding approximately to the integrated luminosity collected by the Belle experiment. The input values are set to $f^{00} = 0.486$ and $f^{+-} = 0.511$. For each pseudo-experiment, the yields of single- and double-tag events are determined and used to perform a fit to the system of equations described in Sec.~\ref{equations}.

To account for contributions from $e^+ e^- \to c \bar{c}$ production, combinatorial background, and the uncertainty associated with the subtraction of scaled off-resonance data, the statistical uncertainty of the single $D^0$ yield is increased by a factor of 8 relative to the naive $\sqrt{N}$ expectation. This factor is estimated using the statistical uncertainty of the inclusive $D^0$ yield and its momentum spectrum reported in Ref.~\cite{Belle:2023yfw}, and also reflects the approximately ten times smaller size of the continuum data sample compared to the on-resonance data. Conservatively, the same factor is applied to the yields $N_{D^0 \bar{D}^0}$ and $N_{D^0 D^0}$.

For $D^{*+}$ mesons, the momentum spectrum in $e^+ e^- \to c \bar{c}$ events is assumed to follow that of $D^0$ mesons, while the combinatorial background contribution is neglected; in this case, the statistical uncertainty $\sqrt{N}$ is scaled by a factor of 2.5. For single-lepton yields, the uncertainty associated with the subtraction of continuum background is estimated using the lepton momentum spectrum in $e^+e^- \to q\bar{q}$ ($q=u,d,s,c$) events reported in Ref.~\cite{BaBar:2004bij}, resulting in an overall scaling factor of 2. For all other observables, the statistical uncertainty is taken to be $\sqrt{N}$.

\begin{figure}
\centering
\includegraphics[width=0.48\linewidth]{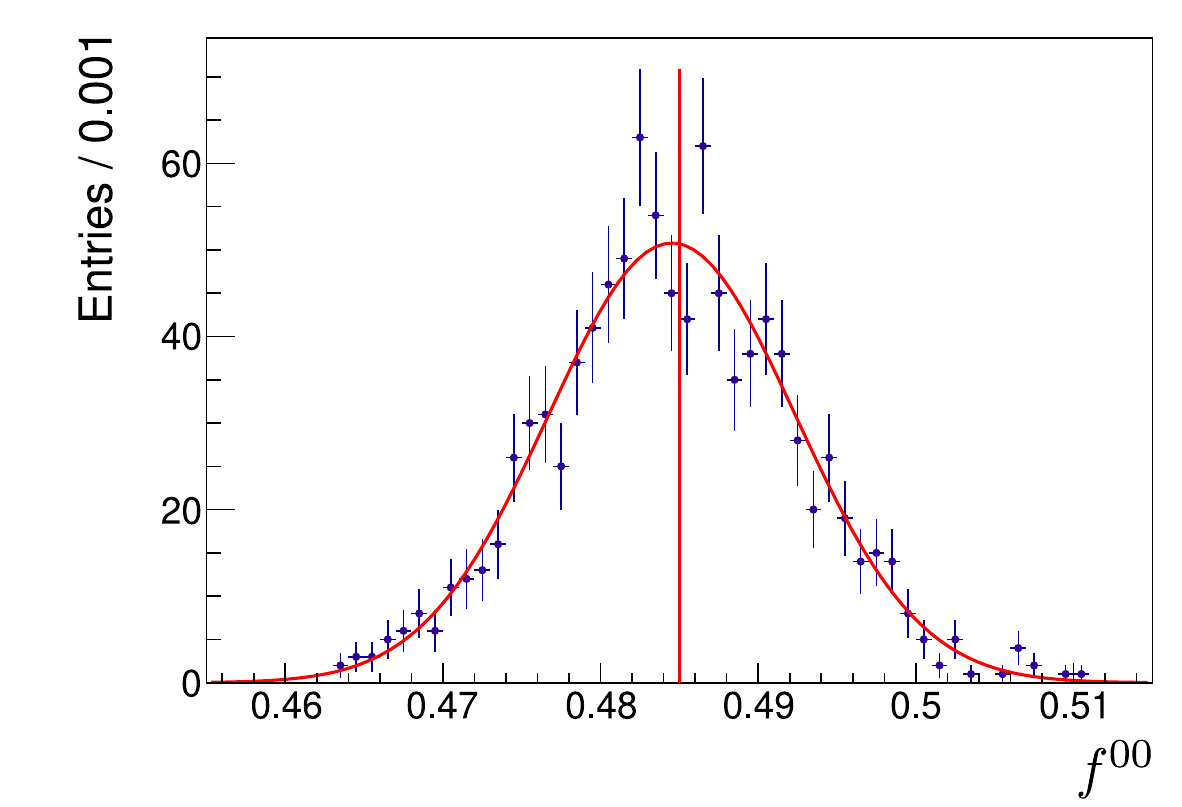}
\includegraphics[width=0.48\linewidth]{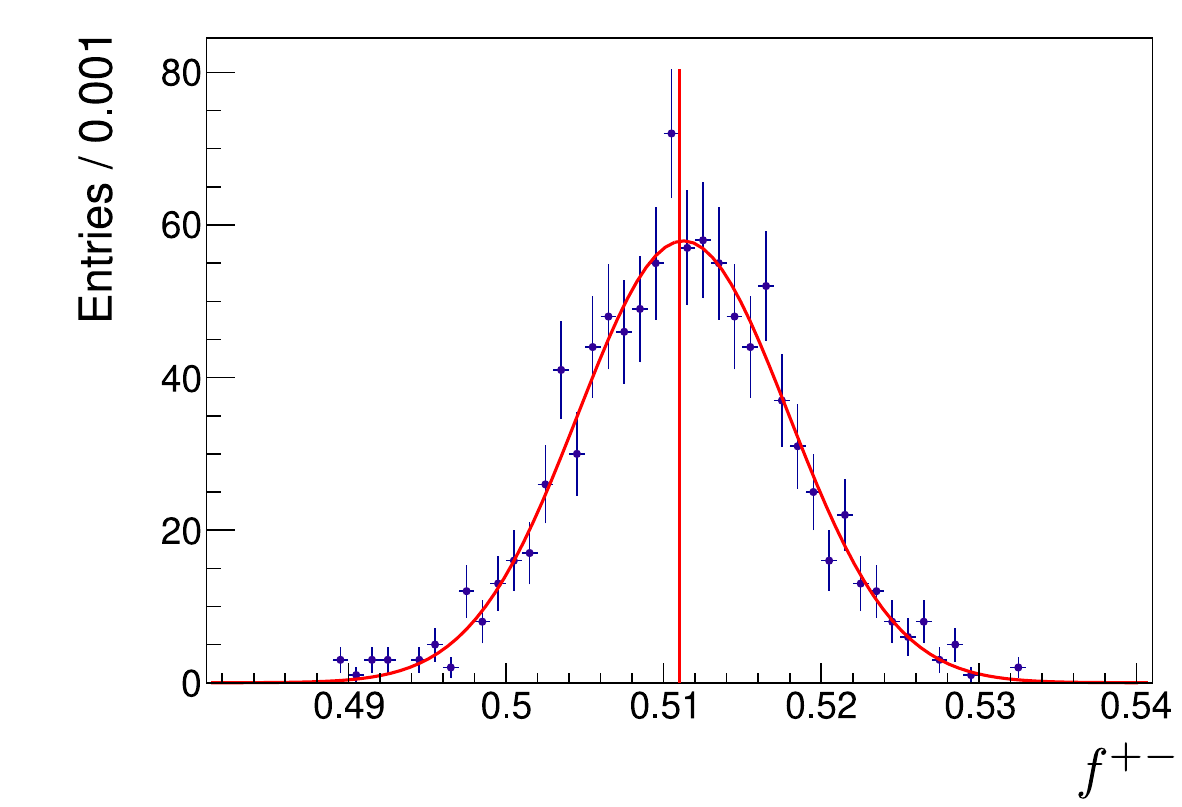}
\caption{Distributions of the fitted values of $f^{00}$ (left) and $f^{+-}$ (right) obtained from pseudo-experiments. The vertical line indicates the input value, while the curve shows the result of a Gaussian fit.}
\label{fig1}
\end{figure}

The distributions of the fitted values of $f^{00}$ and $f^{+-}$ obtained from the pseudo-experiments are shown in Fig.~\ref{fig1}. Both distributions are well described by Gaussian functions, with mean values consistent with the corresponding input parameters. The widths of these distributions provide an estimate of the statistical precision achievable with the proposed method:
\begin{align}
f^{00} &= 0.4845 \pm 0.0079, \\
f^{+-} &= 0.5111 \pm 0.0069.
\end{align}
These results indicate that the proposed method allows the $B\bar{B}$ production fractions at the $\Upsilon(4S)$ resonance to be measured with a precision comparable to the current world-average uncertainty.

An additional feature of the method is its sensitivity to the fraction $f_{\cancel{B}}$ of $\Upsilon(4S)$ decays into non-$B\bar{B}$ final states. In previous measurements, this quantity was determined as the sum of observed exclusive channels and then used in a simultaneous fit with $f^{00}$ and $f^{+-}$ under the constraint $f^{00} + f^{+-} + f_{\cancel{B}} = 1$. The current world-average value is $f_{\cancel{B}} = (0.264^{+1.25}_{-0.02})\%$~\cite{hflav:2024ctg}.

Within the present approach, this fraction can be extracted directly from the fit. The corresponding precision, estimated from pseudo-experiments, is
\begin{align}
f_{\cancel{B}} &= 0.0043 \pm 0.0022.
\end{align}
Such a measurement may provide sensitivity to yet unobserved decay channels of the $\Upsilon(4S)$.

\begin{figure}
\centering
\includegraphics[width=0.48\linewidth]{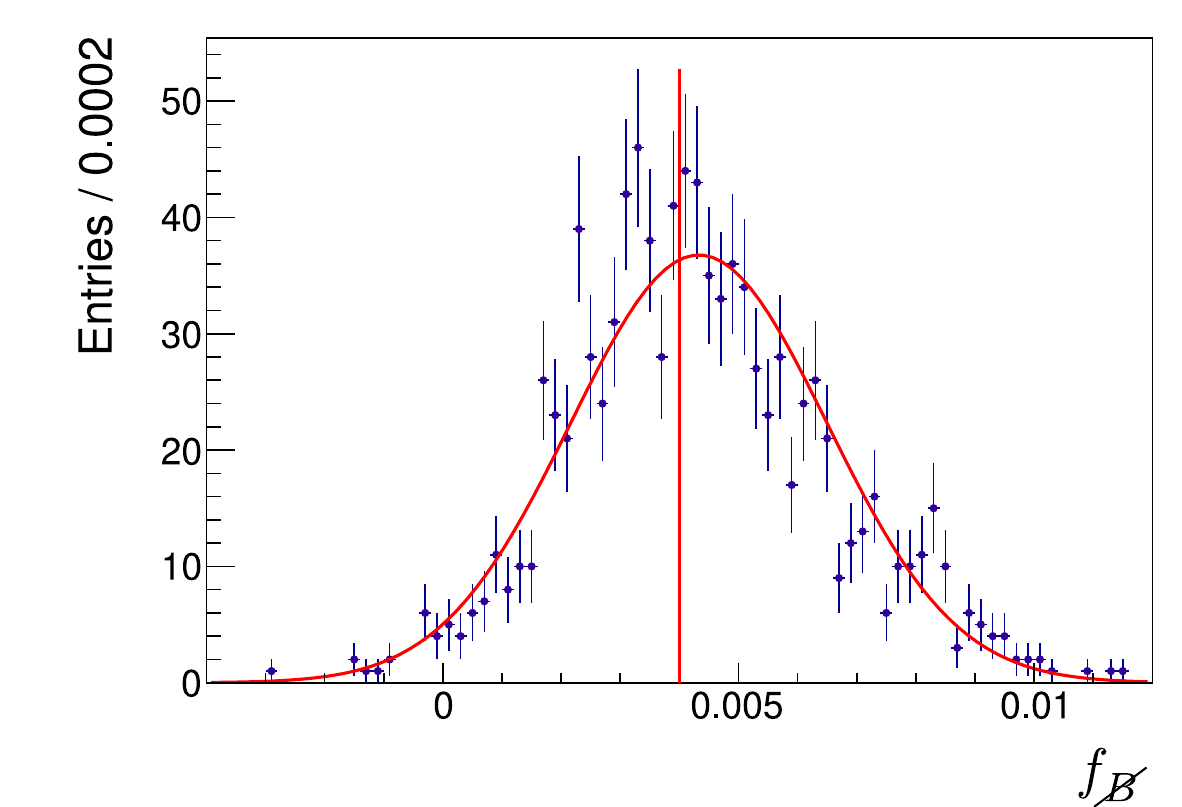}
\caption{Distribution of the fitted values of $f^{\cancel{B}}$ obtained from pseudo-experiments. The vertical line indicates the input value, while the curve shows the result of a Gaussian fit.}
\label{fig2}
\end{figure}

\section{Systematic uncertainties}

The proposed method relies on several simplifying assumptions, whose impact on the extracted parameters should be assessed.

First, in the derivation of the system of equations, it is assumed that charm mesons originating from two different $B$ mesons are emitted into the same hemisphere with a probability of $1/2$. This relation would hold exactly if the $B$ mesons were produced at rest in the $\Upsilon(4S)$ frame. In reality, the $B$ mesons carry a small momentum, which may lead to deviations from this value.

This assumption can be tested directly using data. In particular, one can compare the yields of same-sign $D$-meson pairs reconstructed in the same and in opposite hemispheres. Such pairs are dominantly produced in decays of different $B$ mesons, and therefore provide a clean probe of the underlying angular correlation. This allows the effective value of the same-hemisphere probability to be validated and, if necessary, corrected.

Second, the method assumes that same-sign lepton pairs and $D \ell^+$ combinations originate from different $B$ mesons. In practice, this assumption can be violated by several sources, including hadron misidentification, decays of charged pions and kaons in flight, and photon conversions. As a result, configurations such as $\ell^+ \ell^+$ or $D \ell^+$ may occasionally arise from the decay of a single $B$ meson. The probability of such contributions is expected to be small due to the requirement of high lepton momentum.

Their impact can nevertheless be estimated within the framework of the method by extending the system of equations to include observables corresponding to nominally forbidden combinations, such as $\ell^+ \ell^+ \ell^+$, $\ell^+ \ell^+ \ell^+ \ell^+$, or $D \ell^+ \ell^+$. This extension can be accompanied by the introduction of effective branching fractions for processes such as $B \to \ell^+ \ell^+ X$ and $B \to D \ell^+ \ell^+ X$, which absorb contributions from misidentified or secondary leptons.

In addition, contributions from doubly Cabibbo-suppressed decays, such as $D^0 \to K^+ \pi^-$, can lead to apparent wrong-sign charm assignments. These effects are expected to be small, at the level of approximately $0.4\%$ of the Cabibbo-favored rate, and can be accounted for using known ratios of decay branching fractions.
\label{Systematics}
\section{Summary}

In summary, we propose a new method for measuring the relative production fractions of charged and neutral $B$-meson pairs, $f^{+-}$ and $f^{00}$, in $\Upsilon(4S)$ decays. The method is based on statistical tagging using inclusive production of energetic $D$ mesons and charged leptons, and exploits single- and multi-tag yields to construct a system of equations relating the observables to the underlying production and decay probabilities. The approach does not rely on exclusive reconstruction, assumptions about isospin symmetry, or detailed modelling of decay dynamics. In particular, all relevant branching fractions and reconstruction efficiencies enter as free parameters of the fit and are determined directly from data, without the use of external inputs. As a result, the method provides a fully data-driven and largely model-independent determination of the production fractions.

A feasibility study based on toy Monte-Carlo simulations indicates that, for a dataset corresponding to the full Belle statistics, the expected statistical precision on $f^{00}$ and $f^{+-}$ is at the level of $\mathcal{O}(5\times10^{-3})$. This precision is comparable to the current world-average uncertainty, while being achieved within a single measurement.

The proposed method thus offers a complementary approach to existing techniques, providing the first largely model-independent determination of the $B\bar{B}$ production fractions at the $\Upsilon(4S)$ resonance with competitive precision.

\section*{Acknowledgments\label{sec:acknow}}
The authors would like to thank Roman Mizuk, Alex Bondar and Alexey Drutskoy for useful discussions. This work was funded by the Russian Science Foundation (project No. 25-12-00160).

\newpage
\bibliography{bibl}
\begin{appendices}
\section{Full system of equations}
\label{appendixA}
\subsection{Single-daughter yields}
\begin{eqnarray}
\begin{split}
N_{D^{*+}} = 2N_{\Upsilon(4S)}\Big(f^{00}\big(\Br(B^0 \to D^{*+}X)+\Br(B^0 \to D^{*-}X)\big)\\
+ f^{+-}\big(\Br(B^+ \to D^{*+} X)+\Br(B^+ \to D^{*-} X)\big)\Big)
\end{split}
\end{eqnarray}

\begin{eqnarray}
\begin{split}
N_{D^{0}} = 2N_{\Upsilon(4S)}\Big(f^{00}\big(\Br(B^0 \to D^{0}X) +\Br(B^0 \to \bar{D}^{0}X)\big)\\
+ f^{+-}\big(\Br(B^+ \to D^0 X) + \Br(B^+ \to \bar{D}^0 X)\big)\Big)
\end{split}
\end{eqnarray}

\begin{eqnarray}
\begin{split}
N_{\ell^+} = 2N_{\Upsilon(4S)}\Big(f^{00}\big(\Br(B^0 \to \ell^+ X) +\Br(B^0 \to \ell^- X)\big)\\
+ f^{+-}\big(\Br(B^+ \to \ell^+ X) + \Br(B^+ \to \ell^- X)\big)\Big)
\end{split}
\end{eqnarray}

\subsection{Double-daughter yields}
\begin{eqnarray}
\begin{split}
N_{D^{*+}D^{*+}}= 2N_{\Upsilon(4S)}\Bigg(f^{00}\Big((1-\chi_d)\Br(B^0 \to D^{*+}X)\Br(\bar{B}^0 \to D^{*+}X)\\
+ \frac{\chi_d}{2}\big(\Br(B^0 \to D^{*+}X)^2 + \Br(\bar{B}^0 \to D^{*+}X)^2\big)\Big)\\
+ f^{+-}\Br(B^+ \to D^{*+} X)\Br(B^+ \to D^{*-} X)\Bigg)
\end{split}
\end{eqnarray}

\begin{eqnarray}
\begin{split}
N_{D^{*+}D^{*-}} = \frac{1}{2}N_{\Upsilon(4S)}\Bigg(f^{00}\Big((1-\chi_d)\big(\Br(B^0 \to D^{*+}X)^2+\Br(B^0 \to D^{*-}X)^2\big)\\
+ 2\chi_d\Br(B^0 \to D^{*+}X)\Br(B^0 \to D^{*-}X)\Big)\\
+ f^{+-}\Big(\Br(B^+ \to D^{*+} X)^2 + \Br(B^+ \to D^{*-} X)^2\Big)\Bigg)
\end{split}
\end{eqnarray}

\begin{eqnarray}
\begin{split}
N_{D^{*+}D^{0}}= 2N_{\Upsilon(4S)}\Bigg(f^{00}\Big((1-\chi_d)\Br(B^0 \to D^{*+}X)\Br(\bar{B}^0 \to D^{0}X)\\
+ (1-\chi_d)\Br(B^0 \to D^{0}X)\Br(\bar{B}^0 \to D^{*+}X)\\
+ \chi_d\big(\Br(B^0 \to D^{*+}X)\Br(B^0 \to D^{0}X) + \Br(B^0 \to D^{*-}X)\Br(B^0 \to \bar{D}^{0}X)\big)\Big)\\
+ f^{+-}\Big(\Br(B^+ \to D^{*+} X)\Br(B^- \to D^0 X) + \Br(B^+ \to D^0 X)\Br(B^- \to D^{*+} X)\Big)\Bigg)
\end{split}
\end{eqnarray}

\begin{eqnarray}
\begin{split}
N_{D^{*+} \bar{D}^{0}} = N_{\Upsilon(4S)} \Bigg(f^{00}\Big((1-\chi_d)\Br(B^0 \to D^{*+}X)\Br(\bar{B}^0 \to \bar{D}^{0}X) \\
+(1-\chi_d)\Br(B^0 \to \bar{D}^{0}X)\Br(\bar{B}^0 \to D^{*+}X)\\
+ \chi_d\big(\Br(B^0 \to D^{*+}X)\Br(B^0 \to \bar{D}^{0}X)+\Br(B^0 \to D^{*-}X)\Br(B^0 \to D^{0}X)\big)\Big)\\
+ f^{+-}\Big(\Br(B^+ \to D^{*+}X)\Br(B^- \to \bar{D}^{0}X)+\Br(B^+ \to \bar{D}^{0}X)\Br(B^- \to D^{*+}X)\Big)\Bigg)
\end{split}
\end{eqnarray}
\begin{eqnarray}
\begin{split}
N_{D^{0}D^{0}}= 2N_{\Upsilon(4S)}\Bigg(f^{00}\Big((1-\chi_d)\Br(B^0 \to D^{0}X)\Br(\bar{B}^0 \to D^{0}X)\\
+ \frac{\chi_d}{2}\big(\Br(B^0 \to D^{0}X)^2 + \Br(\bar{B}^0 \to D^{0}X)^2\big)\Big)\\
+ f^{+-}\Br(B^+ \to D^{0} X)\Br(B^+ \to D^{0} X)\Bigg)
\end{split}
\end{eqnarray}

\begin{eqnarray}
\begin{split}
N_{D^{0}  \bar{D}^{0}} = \frac{1}{2}N_{\Upsilon(4S)}\Bigg(f^{00}\Big((1-\chi_d)\big(\Br(B^0 \to D^{0}X)^2+\Br(B^0 \to \bar{D}^{0}X)^2\big)\\
+ 2\chi_d\Br(B^0 \to D^{0}X)\Br(B^0 \to \bar{D}^{0}X)\Big)\\
+ f^{+-}\Big(\Br(B^+ \to D^{0} X)^2 + \Br(B^+ \to \bar{D}^{0} X)^2\Big)\Bigg)
\end{split}
\end{eqnarray}

\begin{eqnarray}
\begin{split}
N_{D^{*+}\ell^+}= 2N_{\Upsilon(4S)}\Bigg(f^{00}\Big((1-\chi_d)\Br(B^0 \to D^{*+}X)\Br(\bar{B}^0 \to \ell^+X)\\
+ (1-\chi_d)\Br(B^0 \to \ell^+X)\Br(\bar{B}^0 \to D^{*+}X)\\
+ \chi_d\big(\Br(B^0 \to D^{*+}X)\Br(B^0 \to \ell^+X) + \Br(B^0 \to D^{*-}X)\Br(B^0 \to \ell^-X)\big)\Big)\\
+ f^{+-}\Big(\Br(B^+ \to D^{*+} X)\Br(B^- \to \ell^+ X) + \Br(B^+ \to \ell^+ X)\Br(B^- \to D^{*+} X)\Big)\Bigg)
\end{split}
\end{eqnarray}

\begin{eqnarray}
\begin{split}
N_{D^{*+} \ell^-} = N_{\Upsilon(4S)} \Bigg(f^{00}\Big((1-\chi_d)\Br(B^0 \to D^{*+}X)\Br(\bar{B}^0 \to \ell^- X)\\
+(1-\chi_d)\Br(B^0 \to \ell^-X)\Br(\bar{B}^0 \to D^{*+}X)\\
+ \chi_d\big(\Br(B^0 \to D^{*+}X)\Br(B^0 \to \ell^- X) + \Br(B^0 \to D^{*-}X)\Br(B^0 \to \ell^+X)\big)\Big)\\
+ f^{+-}\Big(\Br(B^+ \to D^{*+}X)\Br(B^- \to \ell^-X)+\Br(B^+ \to \ell^- X)\Br(B^- \to D^{*+}X)\Big)\Bigg)\\
+2N_{\Upsilon(4S)}\Big(f^{00}\Br(B^0 \to D^{*\mp} \ell^{\pm}X) + f^{+-}\Br(B^+ \to D^{*\mp}\ell^{\pm}X)\Big) 
\end{split}
\end{eqnarray}

\begin{eqnarray}
\begin{split}
N_{D^{0}\ell^+}= 2N_{\Upsilon(4S)}\Bigg(f^{00}\Big((1-\chi_d)\Br(B^0 \to D^{0}X)\Br(\bar{B}^0 \to \ell^+X)\\
+ (1-\chi_d)\Br(B^0 \to \ell^+X)\Br(\bar{B}^0 \to D^{0}X)\\
+ \chi_d\big(\Br(B^0 \to D^{0}X)\Br(B^0 \to \ell^+X) + \Br(B^0 \to \bar{D}^{0}X)\Br(B^0 \to \ell^-X)\big)\Big)\\
+ f^{+-}\Big(\Br(B^+ \to D^{0} X)\Br(B^- \to \ell^+ X) + \Br(B^+ \to \ell^+ X)\Br(B^- \to D^{0} X)\Big)\Bigg)
\end{split}
\end{eqnarray}

\begin{eqnarray}
\begin{split}
N_{D^{0} \ell^-} = N_{\Upsilon(4S)} \Bigg(f^{00}\Big((1-\chi_d)\Br(B^0 \to D^{0}X)\Br(\bar{B}^0 \to \ell^- X)\\
+(1-\chi_d)\Br(B^0 \to \ell^-X)\Br(\bar{B}^0 \to D^{0}X)\\
+ \chi_d\big(\Br(B^0 \to D^{0}X)\Br(B^0 \to \ell^- X) + \Br(B^0 \to \bar{D}^{0}X)\Br(B^0 \to \ell^+X)\big)\Big)\\
+ f^{+-}\Big(\Br(B^+ \to D^{0}X)\Br(B^- \to \ell^-X)+\Br(B^+ \to \ell^- X)\Br(B^- \to D^{0}X)\Big)\Bigg)\\
+2N_{\Upsilon(4S)}\Big(f^{00}\Br(B^0 \to \bar{D}^0/D^{0} \ell^{\pm}X) + f^{+-}\Br(B^+ \to \bar{D}^0/D^{0} \ell^{\pm}X)\Big)
\end{split}
\end{eqnarray}

\begin{eqnarray}
\begin{split}
N_{\ell^+\ell^+}= 2N_{\Upsilon(4S)}\Bigg(f^{00}\Big((1-\chi_d)\Br(B^0 \to \ell^+X)\Br(\bar{B}^0 \to \ell^+X)\\
+ \frac{\chi_d}{2}\big(\Br(B^0 \to \ell^+X)^2 + \Br(\bar{B}^0 \to \ell^+X)^2\big)\Big)\\
+ f^{+-}\Br(B^+ \to \ell^+ X)\Br(B^+ \to \ell^- X)\Bigg)
\end{split}
\end{eqnarray}

\begin{eqnarray}
\begin{split}
N_{\ell^+\ell^-} = N_{\Upsilon(4S)} \Bigg(f^{00}\Big((1-\chi_d)\big(\Br(B^0 \to \ell^+X)^2+\Br(B^0 \to \ell^-X)^2\big)\\
+ 2\chi_d\Br(B^0 \to \ell^+X)\Br(B^0 \to \ell^-X)\Big)\\
+ f^{+-}\Big(\Br(B^+ \to \ell^+ X)^2 + \Br(B^+ \to \ell^- X)^2\Big)\Bigg)\\
+2N_{\Upsilon(4S)}\Big(f^{00}\Br(B^0\to \ell^+ \ell^-X) + f^{+-}\Br(B^+ \to \ell^+ \ell^-X)\Big)
\end{split}
\end{eqnarray}

\subsection{Triple-daughters yields}
\begin{eqnarray}
\begin{split}
N_{D^{*}D^{*}\ell} = N_{\Upsilon(4S)} \Big(f^{00}\big(\Br(B^0 \to D^{*+}X)+\Br(B^0 \to D^{*-}X)\big)\Br(B^0 \to D^{*\mp}\ell^{\pm} X)\\
+ f^{+-}\big(\Br(B^+ \to D^{*+} X)+\Br(B^+ \to D^{*-} X)\big)\Br(B^+ \to D^{*\mp}\ell^{\pm} X))\Big)
\end{split}
\end{eqnarray}

\begin{eqnarray}
\begin{split}
N_{D^{0}D^{*}\ell} = N_{\Upsilon(4S)} \Big(f^{00}\big(\Br(B^0 \to D^{*+}X)+\Br(B^0 \to D^{*-}X)\big)\Br(B^0 \to \DDbar \ell^{\pm} X)\\
+ f^{00}\big(\Br(B^0 \to D^{0}X)+\Br(B^0 \to \bar{D}^{0}X)\big)\Br(B^0 \to D^{*\mp}\ell^{\pm} X)\\
+ f^{+-}\big(\Br(B^+ \to D^{*+} X)+\Br(B^+ \to D^{*-} X)\big)\Br(B^+ \to \DDbar \ell^{\pm} X)\\
+ f^{+-}\big(\Br(B^+ \to D^0 X)+\Br(B^+ \to \bar{D}^0 X)\big)\Br(B^+ \to D^{*\mp}\ell^{\pm} X)\Big)
\end{split}
\end{eqnarray}

\begin{eqnarray}
\begin{split}
N_{D^{0}D^{0}\ell} = N_{\Upsilon(4S)} \Big(f^{00}\big(\Br(B^0 \to D^{0}X)+\Br(B^0 \to \bar{D}^{0}X)\big)\Br(B^0 \to \DDbar \ell^{\pm} X)\\
+ f^{+-}\big(\Br(B^+ \to D^{0} X)+\Br(B^+ \to \bar{D}^{0} X)\big)\Br(B^+ \to \DDbar \ell^{\pm} X))\Big)
\end{split}
\end{eqnarray}

\begin{eqnarray}
\begin{split}
N_{D^{*}\ell\ell} = 2N_{\Upsilon(4S)} \Big(f^{00}\big(\Br(B^0 \to D^{*+}X)+\Br(B^0 \to D^{*-}X)\big)\Br(B^0 \to \ell^+ \ell^- X)\\+ f^{+-}\big(\Br(B^+ \to D^{*+} X)+\Br(B^+ \to D^{*-} X)\big)\Br(B^+ \to \ell^+ \ell^- X)\Big)\\
2N_{\Upsilon(4S)}\Big(f^{00}\big(\Br(B^0 \to \ell^+ X)+\Br(B^0 \to \ell^- X)\big)\Br(B^0 \to D^{*\mp}\ell^{\pm} X)\\
+ f^{+-}\big(\Br(B^+ \to \ell^+ X)+\Br(B^+ \to \ell^- X)\big)\Br(B^+ \to D^{*\mp}\ell^{\pm} X)\Big)
\end{split}
\end{eqnarray}

\begin{eqnarray}
\begin{split}
N_{D^{0}\ell\ell} = 2N_{\Upsilon(4S)} \Big(f^{00}\big(\Br(B^0 \to D^{0}X)+\Br(B^0 \to \bar{D}^{0}X)\big)\Br(B^0 \to \ell^+ \ell^- X)\\+ f^{+-}\big(\Br(B^+ \to D^{0} X)+\Br(B^+ \to \bar{D}^{0} X)\big)\Br(B^+ \to \ell^+ \ell^- X)\Big)\\
2N_{\Upsilon(4S)}\Big(f^{00}\big(\Br(B^0 \to \ell^+ X)+\Br(B^0 \to \ell^- X)\big)\Br(B^0 \to \DDbar \ell^{\pm} X)\\
+ f^{+-}\big(\Br(B^+ \to \ell^+ X)+\Br(B^+ \to \ell^- X)\big)\Br(B^+ \to \DDbar\ell^{\pm} X)\Big)
\end{split}
\end{eqnarray}

\begin{eqnarray}
\begin{split}
N_{\ell \ell\ell} = 2N_{\Upsilon(4S)} \Big(f^{00}\big(\Br(B^0 \to \ell^+ X)+\Br(B^0 \to \ell^- X)\big)\Br(B^0 \to \ell^+ \ell^- X)\\
+ f^{+-}\big(\Br(B^+ \to \ell^+ X)+\Br(B^+ \to \ell^- X)\big)\Br(B^+ \to \ell^+ \ell^- X)\Big)
\end{split}
\end{eqnarray}

\subsection{Four-daughters yields}
\begin{eqnarray}
\begin{split}
N_{D^{*}D^{*}\ell\ell} = \frac{1}{2}N_{\Upsilon(4S)} \Big(f^{00}\Br(B^0 \to D^{*\mp}\ell^{\pm} X)\Br(B^0 \to D^{*\mp}\ell^{\pm} X)\\
+ f^{+-}\Br(B^+ \to D^{*\mp}\ell^{\pm} X)\Br(B^+ \to D^{*\mp}\ell^{\pm} X)\Big)
\end{split}
\end{eqnarray}

\begin{eqnarray}
\begin{split}
N_{D^{*}D^{0}\ell\ell} = N_{\Upsilon(4S)} \Big(f^{00}\Br(B^0 \to D^{*\mp}\ell^{\pm} X)\Br(B^0 \to \DDbar \ell^{\pm} X)\\
+ f^{+-}\Br(B^+ \to D^{*\mp}\ell^{\pm} X)\Br(B^+ \to \DDbar \ell^{\pm} X)\Big)
\end{split}
\end{eqnarray}

\begin{eqnarray}
\begin{split}
N_{D^{0}D^{0}\ell\ell} = \frac{1}{2}N_{\Upsilon(4S)} \Big(f^{00}\Br(B^0 \to \DDbar \ell^{\pm} X)\Br(B^0 \to \DDbar\ell^{\pm} X)\\
+ f^{+-}\Br(B^+ \to \DDbar\ell^{\pm} X)\Br(B^+ \to \DDbar \ell^{\pm} X)\Big)
\end{split}
\end{eqnarray}

\begin{eqnarray}
\begin{split}
N_{D^{*}\ell\ell\ell} = 2N_{\Upsilon(4S)} \Big(f^{00}\Br(B^0 \to D^{*\mp}\ell^{\pm} X)\Br(B^0 \to \ell^+ \ell^- X)\\
+ f^{+-}\Br(B^+ \to D^{*\mp}\ell^{\pm} X)\Br(B^+ \to \ell^+ \ell^- X)\Big)
\end{split}
\end{eqnarray}

\begin{eqnarray}
\begin{split}
N_{D^{0}\ell\ell\ell} = 2N_{\Upsilon(4S)} \Big(f^{00}\Br(B^0 \to \DDbar\ell^{\pm} X)\Br(B^0 \to \ell^+ \ell^- X)\\
+ f^{+-}\Br(B^+ \to \DDbar \ell^{\pm} X)\Br(B^+ \to \ell^+ \ell^- X)\Big)
\end{split}
\end{eqnarray}

\begin{eqnarray}
\begin{split}
N_{\ell\ell\ell\ell} = N_{\Upsilon(4S)} \Big(f^{00}\Br(B^0 \to \ell^+ \ell^- X)\Br(B^0 \to \ell^+ \ell^- X)\\+ f^{+-}\Br(B^+ \to \ell^+ \ell^- X)\Br(B^+ \to \ell^+ \ell^- X)\Big)
\end{split}
\end{eqnarray}

\end{appendices}
\end{document}